\documentclass{ws-ijmpcs}

\newcommand {\apgt} {\ {\raise-.5ex\hbox{$\buildrel>\over\sim$}}\ }
\newcommand {\aplt} {\ {\raise-.5ex\hbox{$\buildrel<\over\sim$}}\ }
\begin{document}

\markboth{AGUDO ET AL.}
{MAPCAT}

\catchline{}{}{}{}{}
%
%%%%%%%%%%%%%%%%%%%%%%%%%%%%%%%%%%%%%%%%%%%%%%%%%%%%%%%%%%%%%%%%%%%%

\title{MAPCAT: MONITORING AGN WITH POLARIMETRY AT THE CALAR ALTO TELESCOPES}

\author{IV\'AN AGUDO}

\address{Instituto de Astrof\'{\i}sica de Andaluc\'{\i}a (CSIC),  Apartado 3004, E-18080 Granada, Spain, and \\
Institute for Astrophysical Research, Boston University, 725 Commonwealth Avenue, Boston, MA 02215, USA; iagudo@iaa.es}

\author{SOL N. MOLINA, JOS\'E L. G\'OMEZ}

\address{Instituto de Astrof\'{\i}sica de Andaluc\'{\i}a (CSIC),  Apartado 3004, E-18080 Granada, Spain}

\author{ALAN P. MARSCHER, SVETLANA G. JORSTAD}

\address{Institute for Astrophysical Research, Boston University, 725 Commonwealth Avenue, Boston, MA 02215, USA}

\author{JOCHEN HEIDT}

\address{ZAH, Landessternwarte Heidelberg, K\"onigstuhl, 69117 Heidelberg, Germany}

\maketitle

\begin{history}
\received{Day Month Year}
\revised{Day Month Year}
\end{history}

\begin{abstract}
We introduce MAPCAT, a long-term observing program for ``Monitoring of AGN with Polarimetry at the Calar Alto Telescopes''. 
Multi-spectral-range studies are critical to understand some of the most relevant current problems of high energy astrophysics of blazars such as their high energy emission mechanisms and the location of their $\gamma$-ray emission region through event associations across the spectrum. 
Adding multi-spectral-range polarimetry allows for even more reliable identification of polarized flares across the spectrum in these kind of objects, as well as for more accurate modeling of their magnetic field. 
As part of a major international effort to study the long term multi-spectral range polarimetric behavior of blazars, MAPCAT uses -since mid 2007- CAFOS on the 2.2m Telescope at the Calar Alto Observatory (Almer\'ia, Spain) to
obtain monthly optical ($R$-band) photo-polarimetric measurements of a sample of 34 of the brightest $\gamma$-ray, optical, and radio-millimeter blazars accessible from the northern hemisphere.
\end{abstract}

\ccode{PACS numbers: 11.25.Hf, 123.1K}

\section{Why a Program like MAPCAT?}	
AGN have the ability to emit profusely from radio wavelengths to $\gamma$-rays. 
Recent multi-waveband observations of AGN indicate that the emission regions over the entire spectrum are related but not exactly co-spatial. 
The linear polarization at mm and optical wavelengths often has similar electric vector position angles (EVPA) and can track each other as they vary, which provides a robust method for locating regions that emit at these three wavebands. 
As demonstrated in previous work\cite{Mar08}\cdash\cite{Agu11b}, if simultaneous optical and mm polarization variations track each other, we can link the variable optical emission with a specific feature on 7\,mm Very Long Baseline Array (VLBA) images, on which different jet features can be resolved with ultra-high angular
resolution ($\sim0.15$ milliarcsecond). 
At X- and $\gamma$-rays, we do not benefit from polarization measurements, but observations often show correlation between high-amplitude flux variations in the mm, optical, X-ray and $\gamma$-ray ranges with different time delays\cite{Larionov:2008p338}. 
The ability to connect high-energy and lower frequency flares provides the connection of optical and mm wave emission regions to sites of X and $\gamma$ radiation. 
The optical-mm connection can be established by polarization variability, to which the location of the high-energy emission can be bootstrapped through correlation of high-energy and lower frequency light curves.

Hence, the sites where the bulk of the X-ray, $\gamma$-ray, and optical radiation is emitted can be connected to the positions of jet features (resolved by the VLBA) through correlation of polarization and total flux curves across the spectrum.
Optical polarimetry is here an essential ingredient to link high-energy flares with strong VLBA jet features identifiable through common optical and mm properties.

\section{The Program}	
MAPCAT\footnote{\tt http://www.iaa.es/$\sim$iagudo/research/MAPCAT/} is a long term monitoring program, conducted at the Calar Alto 2.2\,m Telescope, using CAFOS (the Calar Alto Faint Object Spectrograph) in polarimetric imaging mode. 
MAPCAT is designed to study the time evolution of the optical ($R$-band) polarimetric properties of a set of 34 gamma-ray bright blazars: 3C~66A, AO~0235+16, CTA~26, 3C~111, PKS~0420$-$01, PKS~0528+134, S5~0716+71, PKS~0735+17, OJ248, OJ049, 4C~71.07, OJ287, S4~0954+65, PKS~1055+0181, Mrk~421, PKS~1127$-$145, 4C~29.45, ON~231, PKS~1222+216, 3C~273, 3C~279, B2~1308+30, PKS~1406$-$076, PKS~1510$-$08, DA~406, PKS~1622$-$29, 4C~38.41, 3C~345, NRAO~530, OT~081, BL~Lac, 3C~446, CTA~102, 3C~454.3. These are among the brightest $\gamma$-ray AGN, most of them blazars, with $m_{V}\aplt18$, a requirement for us to measure optical polarization optimally- and with declinations accessible to the Calar Alto and other relevant northern observatories. 
These 34 sources are also required to be bright in radio and mm-wavelengths to allow efficient 7mm VLBA imaging and radio-to-mm single dish polarimetric monitoring.

Monitoring the blazar emission in the high energy domain (especially during flares) is key to understand the entire picture of jet emission. 
The major blazar outbursts that we use to look for correlations, from the mm to the $\gamma$-ray regimes, may have time-scales longer than a year, and they are (in general) unpredictable. 
Instead of choosing a single source, we need to monitor as statistically significant sample as possible.
This helps to improve the chances to catch clear flares to find robust multi spectral range correlations.
Our observations are performed once every month in service mode, which at the Calar Alto Observatory ensures as adequate as possible weather conditions. 
The program started mid 2007 (when the Fermi Gamma-Ray Space Telescope was planned to be launched) and it is
expected that it will be active until the end of the operations of Fermi, at least until mid 2013.

\section{MAPCAT as Part of Larger Programs}
MAPCAT contributes to major multi-spectral-range campaigns organized by the WEBT (Whole Earth Blazar Telescope) and GASP (GLAST-AGILE support program) consortia, that are aimed at studying key aspects of the physics of blazars.\cite{Rai08a}\cdash\cite{Pal11} 
Other good examples of these studies are those led by the Fermi-LAT\cite{Abd10a}\cdash\cite{Sch11}, and AGILE\cite{DAm09}\cdash\cite{Ver11} $\gamma$-ray observatory collaborations.

\begin{figure}[pb]
\centerline{\psfig{file=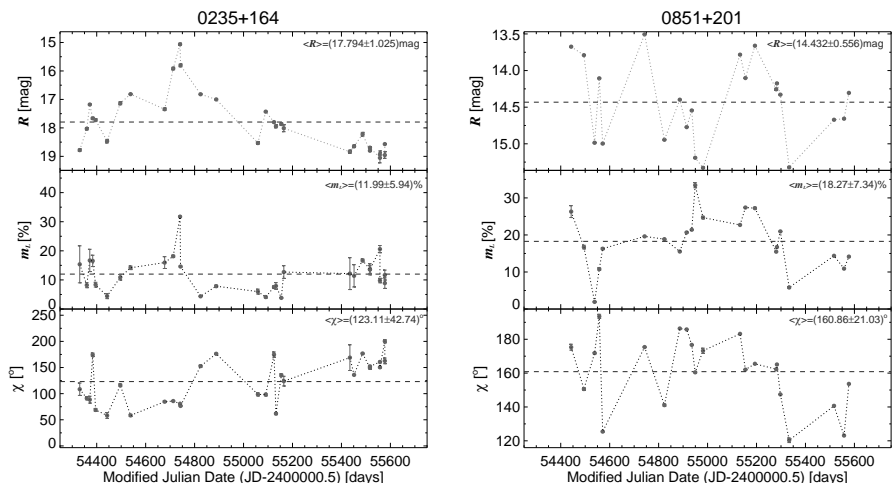,width=12.cm}}
\vspace*{8pt}
\caption{$R$-band evolution curves of the total flux, linear polarization degree, and polarization angle (top to bottom) of AO~0235+164 (left) and OJ287 (right). \label{f1}}
\end{figure}

MAPCAT is also supported by a parallel 3mm \& 1mm polarimetric monitoring program of the MAPCAT source sample conducted at the IRAM 30\,m Telescope, and led by the Relativistic Jet and Blazar Group at the IAA-CSIC. Such program, which observes every $\sim$2 weeks, is performed through the Monitoring AGN with Polarimetry at the IRAM-30m-Telescope (MAPI), and the Polarimetric AGN Monitoring with the IRAM-30m-Telescope (POLAMI) projects.

Both MAPCAT and the IRAM 30\,m programs support a larger scale international multi-spectral-range monitoring for the study of blazars led by the Boston University Blazar Group. 
This program also includes:
(1) Monthly optical polarimetry with the Lowell Obs. 1.8m Perkins Telescope, the St. Petersburg University 0.4m Telescope, and the Crimean Astrophysical Obs. 0.7m Telescope.
Note that overlapping observations of sources reduces the risk of missing measurements due to adverse weather. Moreover, favorable conditions allow short-term variability studies.
(2) Monthly 7\,mm polarimetric VLBA monitoring, and polarimetric Global mm-VLBI Array 3mm observations of 17 sources.
(3) X-ray, UV, and optical flux monitoring up to once per month (or better) with the {\it Swift} X-ray satellite through publicly available data from {\it Swift} programs.
(4) $\gamma$-ray flux monitoring with Fermi. This adds daily fluxes and continuum spectra at $\gamma$-ray energies.
(5) Total flux density light curves in the near-IR and optical at the Liverpool Telescope, and at Lowell Obs., and other telescopes. 
To these we add light curves from ongoing monitoring programs at 1mm at the SMA (through
collaboration with M. Gurwell), plus 8mm at Mets\"ahovi and 2 cm at U. Michigan, groups with which we have long collaborated.

Since the MAPCAT program started, our optical polarimetric data (of which we show some examples in Fig.~\ref{f1}), combined with data from the IRAM 30\,m monitoring programs led by the IAA-CSIC Relativistic Jet and Blazar Group, and those led by the Boston University Blazar Group (1 to 5 above, among others), have significantly contributed to understand the behavior of several sources in our sample.
In particular, our data helped to reveal clear cases of large $R$-band EVPA rotations and correlations with other spectral bands\cite{Mar10,Jor10}, and to unambiguously locate the $\gamma$-ray flaring emission region\cite{Agu11a,Agu11b} in some blazars.

\section{Acknowledgements}
The Calar Alto Observatory is jointly operated by the Max-Planck-Institut f\"ur Astronomie-MPG and the Instituto de Astrof\'isica de Andaluc\'ia-CSIC.

\end{document}